\documentclass{IEEEtaes}

\usepackage{amsmath,graphicx}
\usepackage{xparse}
\usepackage{caption}
\usepackage{subcaption}
\usepackage{epstopdf}  %for graphics                		
\usepackage{amsmath,amssymb}
\usepackage{algorithm}
\usepackage{lineno,hyperref}
\usepackage[noend]{algpseudocode}
\usepackage{textcomp}
\usepackage{cite}
\makeatletter
\newcommand*{\rom}[1]{\expandafter\@slowromancap\romannumeral #1@}
\def\algbackskip{\hskip-\ALG@thistlm}
\makeatother

\jvol{XX}
\jnum{XX}
\jmonth{XXXXX}
\paper{XXXXX}
\pubyear{2024}
\doiinfo{TAES.2024.Doi Number}

\begin{document}

\title{Towards Adaptive Subspace Detection in Heterogeneous Environment}

\author{Aref Miri Rekavandi}
\member{Member, IEEE}
\affil{Faculty of Engineering and Information Technology, The University of Melbourne, Australia.}

%% \author{FOURTH D. AUTHOR}
%% \affil{University of Colorado, Colorado, USA}

%% \accepteddate{XXXXX XX XXXX}
%% \publisheddate{XXXXX XX XXXX}

%\corresp{The name of the corresponding author appears after the financial information, e.g. {\itshape (Corresponding author: M. Smith)}. Here you may also indicate if authors contributed equally or if there are co-first authors.}

\authoraddress{Aref Miri Rekavandi is with Faculty of Engineering and Information Technology, The University of Melbourne, Australia.
(e-mail: \href{aref.mirirekavandi@unimelb.edu.au}{aref.mirirekavandi@unimelb.edu.au}). The Matlab code of the results presented in this paper can be found in \url{https://github.com/arekavandi/Heterogeneous_Detector}. This manuscript was submitted to IEEE TAES. }

%\editor{Mentions of supplemental materials and animal/human rights statements can be included here.}
%\supplementary{Color versions of one or more of the figures in this article are available online at \href{http://ieeexplore.ieee.org}{http://ieeexplore.ieee.org}.}

\markboth{Miri Rekavandi}{Towards Adaptive Subspace Detection in Heterogeneous Environment}
\maketitle

\newcommand{\la}{\left\{}
\newcommand{\ra}{\right\}}
\newcommand{\lc}{\left[}
\newcommand{\rc}{\right]}
\newcommand{\lp}{\left(}
\newcommand{\rp}{\right)}
\newcommand{\bd}{\begin{displaymath}}
\newcommand{\ed}{\end{displaymath}}
\newcommand{\beq}{\begin{equation}}
\newcommand{\eeq}{\end{equation}}
\newcommand{\diag}{\textrm{diag}}
\newcommand{\asinh}{\textrm{asinh}}
\newcommand{\EE}{\EuScript{E}}
\newcommand{\mf}{\mathbf}
\newcommand{\bs}{\boldsymbol}
\newcommand{\ud}{\mathrm{d}}
\newcommand{\bfh}{\textbf{h}}
\newcommand{\bff}{\textbf{f}}
\newcommand{\bfz}{\textbf{z}}
\newcommand{\bfy}{\textbf{y}}
\newcommand{\bfv}{\textbf{v}}
\newcommand{\bfx}{\textbf{x}}
\newcommand{\bfs}{\textbf{s}}
\newcommand{\bfX}{\textbf{X}}
\newcommand{\bfe}{\textbf{e}}
\newcommand{\bfU}{\textbf{U}}
\newcommand{\bfu}{\textbf{u}}
\newcommand{\bfH}{\textbf{H}}
\newcommand{\bfK}{\textbf{K}}
\newcommand{\bfC}{\textbf{C}}
\newcommand{\bfV}{\textbf{V}}
\newcommand{\bfI}{\textbf{I}}
\newcommand{\bfW}{\textbf{W}}
\newcommand{\parrow}{\overset{p}{\rightarrow}}
\newcommand{\darrow}{\overset{d}{\rightarrow}}
\newcommand{\bfzero}{\textbf{0}}
\newcommand{\Larrow}{\overset{L}{\rightarrow}}
\newtheorem{acknowledgement}[theorem]{Acknowledgement}
\newtheorem{axiom}[theorem]{Axiom}
\newtheorem{case}[theorem]{Case}
\newtheorem{claim}[theorem]{Claim}
\newtheorem{conclusion}[theorem]{Conclusion}
\newtheorem{condition}[theorem]{Condition}
\newtheorem{conjecture}[theorem]{Conjecture}
\newtheorem{corollary}[theorem]{Corollary}
\newtheorem{criterion}[theorem]{Criterion}
\newtheorem{definition}[theorem]{Definition}
\newtheorem{example}[theorem]{Example}
\newtheorem{exercise}[theorem]{Exercise}
\newtheorem{notation}[theorem]{Notation}
\newtheorem{problem}[theorem]{Problem}
\newtheorem{proposition}[theorem]{Proposition}
\newtheorem{remark}[theorem]{Remark}
\newtheorem{solution}[theorem]{Solution}
\newtheorem{summary}[theorem]{Summary}

\begin{abstract} In this paper, we aim to take one step forward to the scenario where an adaptive subspace detection framework is required to detect subspace signals in non-stationary environments.
	Despite the fact that this scenario is more realistic, the existing studies in detection theory mostly rely on homogeneous, or partially homogeneous assumptions in the environments for their design
	process meaning that the covariance matrices of primary and secondary datasets are exactly the same or different up to a scale factor. In this study, we allow some partial information of the
	train covariance matrix to be shared with the primary dataset, but the covariance matrix in the primary set can be entirely different in the structure. This is particularly true in radar systems where
	the secondary set is collected in distinct spatial and time zones. We design a Generalized Likelihood Ratio Test (GLRT) based detector where the noise is multivariate Gaussian and the subspace
	interference is assumed to be known. The simulation results reveal the superiority of the proposed approach in comparison with conventional detectors for such a realistic and general scenario.
\end{abstract}

\begin{IEEEkeywords}Adaptive subspace detection, maximum likelihood, non-stationary environment, GLRT.
\end{IEEEkeywords}

\section{Introduction}\label{sec:introduction}
\IEEEPARstart{A}{daptive} detection frameworks are widely used as a final step in various signal processing related applications, in particular radar \cite{coluccia2022design,rekavandi2021robust,liu2017tunable,kraut2001adaptive,sun2021adaptive}, communication systems \cite{baktash2017detection,liu2015modified}, and medical imaging analysis \cite{rekavandi2022adaptive,ardekani1999activation} to examine the presence of a target signal or in general a target subspace in noisy multichannel measurements. Depending on the applications, the noise model can vary from a simple Gaussian noise model \cite{scharf1994matched} to highly complex non-parametric \cite{gul2016robust} or non-Gaussian noise models \cite{rekavandi2023trpast}. Several frameworks are typically used for detecting such signals including Generalized Likelihood Ratio Test (GLRT) \cite{rekavandi2019adaptive,kelly1986adaptive,rekavandi2020robust,raghavan2022performance,rekavandi2020adaptive}, Rao or Score test \cite{rekavandi2022adaptive,de2007new,rekavandi2021robust,liu2015modified,sun2021rao}, Wald test \cite{liu2017tunable,naderpour2015generalized,rekavandi2023extended,liu2018multichannel}, Gradient test \cite{ye2022adaptive,tang2022adaptive,terrell2002gradient} and Durbin test \cite{ye2022adaptive,sun2022multichannel}. These variants might differ due to (i) presence of interference up to known/unknown subspace, (ii) known/unknown signal subspace, (iii) prior information about the signal location in the assumed subspace,  (iv) dependency of the noise to the target signal, (v) structures in covariance matrices, (vi) spread/single target, (vii) single/two-step design, etc. For those readers who are interested in the details of these frameworks we refer to \cite{liu2022multichannel} which gives a comprehensive review of the existing studies in the area of multichannel signal detection. 

For estimation of the noise covariance matrices, a typical assumption of access to a secondary target free observations was made in \cite{kelly1986adaptive}. For simplicity, most studies assumed a shared covariance between the secondary and the primary observations (known as the homogeneous case) or assumed their covariance matrices can be different only up to a scale factor (known as partially homogeneous environment). Even though these assumptions may be adequate for some applications, for a majority of the real world problems the real process is different and these constraints are strong for deriving a reliable result. For example, in radar data acquiring process, this means that the cell under the test is enforced to have similar or different (up to a scale factor) covariance matrix compared to other adjacent cells which can be distant from the test cell with different land/air properties.   

In a recent study \cite{coluccia2022glrt}, the authors for the first time studied the scenario where the power of the secondary set noise samples are different from each other and from the test data to detect point-like targets. Later on \cite{yan2023innovative,yin2023classification} extended this scenario to multiple target detection and multiple hypothesis tests.  In this study, we consider a more realistic case where not only the covariance matrices of the samples in the secondary set can differ in scales (powers), but also the covariance matrices of primary and secondary sets can be entirely different in shape within a defined proximity. To do this, we consider the scenario where the signal has a subspace form and it is contaminated with unknown interference (with known subspace). This new subspace detector can be seen as a new memebr of the new generation of detectors which are adapted to a dynamical environment and can be better adapted to real radar systems. The main contributions of this paper may be summarized as follows:
\begin{itemize}
	\item Developing an approximated GLRT based subspace detection method which relaxes the assumption of similar covariance matrices between the primary and secondary observation sets in radar systems.
	\item The proposed subspace detector also takes into account the interference subspace model and allows the secondary samples to differ from each other in terms of covariance matrix structures up to a scale factor.
	\item Simulation results reveal the superiority and robustness of the proposed detector compared to the classical partially homogeneous-based detector, in particular adaptive matched filter.
\end{itemize}  
The remainder of this paper is organized as follows. Section \ref{sec:problem} briefly formulates the problem. Section \ref{sec:background} contains the background on the Adaptive Subspace Detector (ASD) and Adaptive Matched Filter (AMF). In section \ref{sec:methods}, the proposed method is introduced. The simulation results are presented in section \ref{sec:results} and the results are discussed in section \ref{discussion}. Finally, the paper is concluded in section \ref{sec:conclusion}.\\
\textbf{\textit{Notations:}} In this paper, vectors and matrices are denoted by bold-face lower case and upper-case letters. For a given matrix $\textbf{X}$, $\textbf{x}^i$ and $\textbf{x}_j$ represent the $i-$th row and the $j-$th column of the matrix, respectively. Symbols $\textmd{det}(.)$, $\textmd{tr}(.)$, $\log(.)$, $\exp(.)$, $(.)^\top$, $(.)^\dagger$, and $(.)^{-1}$ denote the determinant,  trace, natural logarithm, exponential function, transpose, conjugate transpose, and inverse, respectively. The notation $\sim$ means "is distributed as," and $\mathcal{C}\mathcal{N}(\textbf{m},\textbf{R})$ denotes distribution of a complex normal random vector with mean $\textbf{m}$ and covariance matrix $\textbf{R}$. $\textbf{P}_{\textbf{D}}=\textbf{D}[\textbf{D}^\dagger\textbf{D}]^{-1}\textbf{D}^\dagger$ is the orthogonal projection on the subspace $\langle\textbf{D}\rangle$ where $\langle\textbf{D}\rangle$ is the subspace spanned by columns of matrix $\textbf{D}$, and $\textbf{P}_{\textbf{D}^\perp}=\textbf{I}-\textbf{P}_{\textbf{D}}$ projects on the orthogonal subspace to $\langle\textbf{D}\rangle$ where $\textbf{I}$ is the identity matrix. $HS(.)$ is the Heaviside step function and $\max(a,b)$ returns the maximum of two values.

%------------------------------------------------------------------
\section{Problem Formulation}
\label{sec:problem}
\begin{figure}[!t] 
	\centering
	\includegraphics[width=1\linewidth]{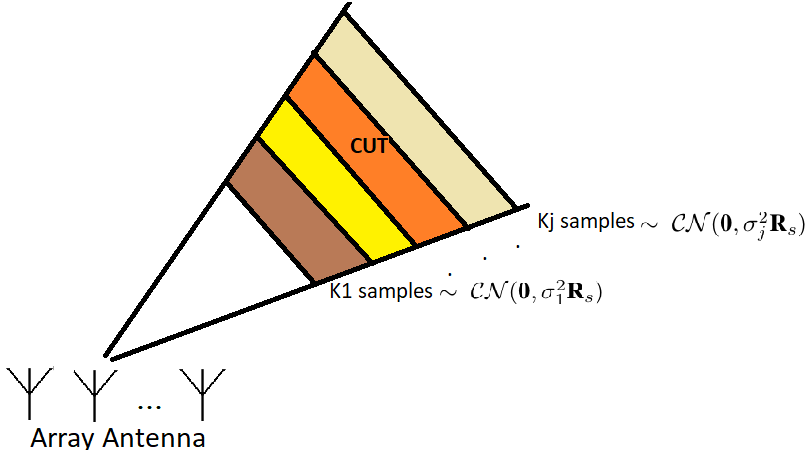}
	\label{f:1a}
	\caption{Visualizing the scenario where the CUT and the Adjacent Cells (ACs) can be generated from different noise distributions. For the $j^{th}$ AC, $K_j$ samples are drawn, and $K_1+K_2+\cdots+K_J=K$.  }
	\label{fig1}
\end{figure}

%------------------------------------------------------------------
Consider an array of $N$ sensors or equivalently a single sensor with $N$ snapshots, which observe signal from a underlying process (e.g., reflected echos received from a target to the radar, recorded magnetic or electrical signals initiated from the neural activation in the brain, etc.). The observation vector (called test data) $\textbf{y} \in \mathbb{C}^N$ often can be decomposed to three main components using a linear model as in \cite{scharf1994matched}
\begin{equation}\label{eq1}
\textbf{y}={\textbf{H}\bs{\theta}}+{\textbf{B}\bs{\phi}}+\boldsymbol{\xi},
\end{equation}
where matrices $\textbf{H}\in \mathbb{C}^{N\times p}$ and $\textbf{B}\in \mathbb{C}^{N\times t}$ are known and linearly independent matrices and their columns form the signal and interference subspace, respectively. In the case of unknown subspace, they might be estimated using Principal Component Analysis (PCA) \cite{rekavandi2020robust,aref2023learning} or Dictionary Learning (DL) techniques \cite{seghouane2023rbdl,khalid2023efficient}.  The target signal can be any linear combination of signal subspace columns where the weights are represented by $\boldsymbol{\theta}$, an unknown deterministic. Similarly, $\boldsymbol{\phi}$ is also unknown deterministic which contributes in the interference part of the test data point. $\boldsymbol{\xi}$ is the noise term which is circularly symmetric complex Gaussian distributed, i.e, $\boldsymbol{\xi} \sim \mathcal{C}\mathcal{N}(0,\sigma^2\textbf{R})$ with unknown covariance matrix $\sigma^2\textbf{R}$. Up to this point, we adopted the same problem setup as the literature in the context of homogeneous detection. Now we assume that there is a target free secondary set, i.e., $\{\textbf{n}_k\}_{k=1}^K$ which partially share the same information about the covariance matrix $\textbf{R}$. This set includes $K$ samples with covariance matrix denoted by $\sigma^2_k\textbf{R}_s$, but each individual covariance matrix is different from the test data (within a predefined proximity) and from the other secondary covariance matrices (up to a scalar). In a particular case, few samples can represent the same environment (in the context of radar, it is called the same adjacent cell) and consequently share the same covariance (and the same scale). This setup id shown in Fig. \ref{fig1} where for each adjacent cell (AC), few samples are drawn and with the total number of $K$ samples.  In other words, it is assumed that $\textbf{n}_k \sim \mathcal{C}\mathcal{N}(\textbf{0},\sigma^2_k\textbf{R}_s)$ where $\sigma^2_k$ and $\textbf{R}_s$ are both unknown. In order to model the non-homogeneity, we assume the covariance of the test data is in the neighborhood of $\textbf{R}_s$ using a measure of proximity, i.e., Frobenius norm, etc. This allows the model to be adequate for the scenario where the environment is non-stationary and the noise characteristics can be different even between each single data collection which makes the model more realistic. Assuming above model, the detection problem now becomes a hypothesis testing task which is a decision between   

\begin{equation}\label{H01}
{\mathcal{H}_0}=\left\{
\begin{array}{@{}l@{\thinspace}l}
\textbf{y}=\textbf{B}\bs{\phi}+\bs{\xi} \sim \mathcal{C}\mathcal{N}(\textbf{B}\bs{\phi},\sigma^2\textbf{R})\\
\textbf{y}_k=\textbf{n}_k \sim \mathcal{C}\mathcal{N}(\textbf{0},\sigma^2_k\textbf{R}_s), k=1,\cdots,K\\
\end{array}
\right.,
\end{equation}
and
\begin{equation}\label{H11}
{\mathcal{H}_1}=\left\{
\begin{array}{@{}l@{\thinspace}l}
\textbf{y}=\textbf{H}\boldsymbol{\theta}+\textbf{B}\bs{\phi}+\bs{\xi} \sim \mathcal{C}\mathcal{N}(\textbf{H}\boldsymbol{\theta}+\textbf{B}\bs{\phi},\sigma^2\textbf{R})\\
\textbf{y}_k=\textbf{n}_k \sim \mathcal{C}\mathcal{N}(\textbf{0},\sigma^2_k\textbf{R}_s), k=1,\cdots,K \\
\end{array}
\right.,
\end{equation}
where $\mathcal{H}_0$ and $\mathcal{H}_1$ are the null and alternative hypotheses, respectively. In above problem, we assume $\|\textbf{R}-\textbf{R}_s\|_F^2\leq \epsilon$, where $\epsilon$ determines the strength of statistical changes between train and test sets, and finally some of the $\textbf{y}_k$s might have the same covariance matrices ($\sigma_j^2\textbf{R}_s$) due to repeated signal collection or having the same statistical behavior for some ACs. 

%------------------------------------------------------------------
\section{Background}
\label{sec:background}
GLRT-based detectors were extensively used in the literature to detect signals which are described by linear model (\ref{eq1}). However, in their design process, it was assumed that the secondary dataset is drawn from a stationary density function (i.e.,  $\mathcal{C}\mathcal{N}(\textbf{0},\textbf{R})$). Using this assumption, sample covariance matrix appears in the final form of detectors. For example in Adaptive Subspace Detector (ASD), the detector takes the form  

\begin{equation}\label{ASD}
\ell_{ASD}(\textbf{y})= \frac{\tilde{\textbf{y}} \textbf{P}_{\tilde{\textbf{B}}^\perp}\textbf{P}_{\tilde{\textbf{H}}}\textbf{P}_{\tilde{\textbf{B}}^\perp} \tilde{\textbf{y}}}{\tilde{\textbf{y}}\textbf{P}_{\tilde{\textbf{B}}^\perp}\tilde{\textbf{y}}},
\end{equation}
where $\tilde{\textbf{H}}={\textbf{S}}^{-\frac{1}{2}} \textbf{H}$, $\tilde{\textbf{B}}={\textbf{S}}^{-\frac{1}{2}} \textbf{B}$, $\tilde{\textbf{y}}=\textbf{S}^{-\frac{1}{2}}\textbf{y}$,  and $\textbf{S}=\frac{1}{K}\sum_{k=1}^{K}\textbf{n}_k\textbf{n}_k^\dagger$ is the sample covariance matrix. ASD can be derived when we maximize the likelihood functions under each hypothesis with respect to all unknown parameters including the covariance matrix. To make the detector computationally efficient, in Adaptive Matched Filter (AMF), it was suggested to initially assume the covariance matrix is known, then in the next step, it can be replaced with the estimated covariance using the secondary set. Finally, AMF takes the form 
\begin{equation}\label{AMF}
\ell_{AMF}(\textbf{y})=\frac{\tilde{\textbf{y}}^\dagger\textbf{P}_{\tilde{\textbf{B}}^\perp}\tilde{\textbf{y}}}{\tilde{\textbf{y}}^\dagger\textbf{P}_{\tilde{\textbf{C}}^\perp}\tilde{\textbf{y}}},
\end{equation} 
where $\tilde{\textbf{C}}=\textbf{S}^{-\frac{1}{2}}\textbf{C}$,  and  $\textbf{C}=[\textbf{H},\textbf{B}]$. Geometrically, it seems that ASD offers a measure of signal energy to signal plus noise energy. In theory, this value is between zero and one, and when it tends to one, it means that we have a significantly large signal energy in the observation and it is sufficient to decide $\mathcal{H}_1$. On the other hand AMF measures the ratio of signal plus noise energy to just energy of the noise. When a significantly large value is observed in the output of the detector, most likely $\mathcal{H}_1$ is correct.  

%------------------------------------------------------------------
\section{Proposed Method}
\label{sec:methods}
For defined problem of subspace detection in non-homogeneous environment, we use a constrained GLRT framework. Assuming a single test data vector $\textbf{y}$ and the secondary data $\{\textbf{n}_k\}_{k=1}^K$ captured in $J$ ACs, the joint likelihood function assuming their independence is given by   
\begin{equation}\label{5}
f\left(\textbf{y},\textbf{n}_1,\dots,\textbf{n}_K\right)= f\left(\textbf{y}\right)\prod_{k_1=1}^{K_1} f\left(\textbf{n}_{k_1}\right)\cdots\prod_{k_J=1}^{K_J} f\left(\textbf{n}_{k_J}\right).
\end{equation}
For a zero mean circularly symmetric complex Gaussian random variable $\boldsymbol{\xi} \sim \mathcal{C}\mathcal{N}(0,\sigma^2\textbf{R})$, the likelihood function is defined by 
\begin{equation}\label{6}
f(\boldsymbol{\xi})= \frac{1}{\pi^N \textmd{det}(\sigma^2\textbf{R})} \exp\{-\frac{1}{\sigma^2}\boldsymbol{\xi}^\dagger \textbf{R}^{-1} \boldsymbol{\xi} \}.
\end{equation} 
Using (\ref{eq1}), (\ref{5}), and (\ref{6}), and by defining  $\bs{\beta}=[\boldsymbol{\theta}^\top, \boldsymbol{\phi}^\top]^\top$, under $\mathcal{H}_1$ we have
\begin{eqnarray}
&&f_1\left(\textbf{y},\textbf{n}_1,\dots,\textbf{n}_K\right)\nonumber\\
&&=\frac{1}{\pi^N \textmd{det}(\sigma^2\textbf{R})}\exp\{-\frac{1}{\sigma^2}(\textbf{y}-{\textbf{C}\bs{\beta}})^\dagger \textbf{R}^{-1} (\textbf{y}-{\textbf{C}\bs{\beta}}) \}\nonumber\\
&&\times\left\{\frac{1}{\pi^N \sigma_1^{2NK_1/K}\cdots \sigma_J^{2NK_J/K}\textmd{det}(\textbf{R}_s)}\right\}^{K}\times\prod_{k_1=1}^{K_1}\nonumber\\
&&\exp\{-\frac{1}{\sigma_1^2}\textbf{n}_{k_1}^\dagger \textbf{R}_s^{-1} \textbf{n}_{k_1}\}\cdots\prod_{k_J=1}^{K_J}\exp\{-\frac{1}{\sigma_J^2}\textbf{n}_{k_J}^\dagger \textbf{R}_s^{-1} \textbf{n}_{k_J}\},\nonumber
\end{eqnarray}
where $J\leq K$ and $K_1+K_2+\cdots+K_J=K$. Under $\mathcal{H}_0$, the density $f_0=f_1|_{\bs{\theta}=\textbf{0}}$. Finally, noting that $\textbf{z}^\dagger\textbf{R}^{-1}\textbf{z}=\textmd{tr}(\textbf{z}^\dagger\textbf{R}^{-1}\textbf{z})=\textmd{tr}(\textbf{R}^{-1}\textbf{z}\textbf{z}^\dagger)$, we can rewrite $f_1$ as
\begin{eqnarray}
&&f_1\left(\textbf{y},\textbf{n}_1,\dots,\textbf{n}_K\right)\nonumber\\
&&=\bigg\{\frac{\sigma_1^{-2NK_1/K+1}\cdots\sigma_J^{-2NK_J/K+1}}{\pi^N \textmd{det}(\textbf{R})^{1/K+1}\textmd{det}(\textbf{R}_s)^{K/K+1}\sigma^{2N/K+1}}\nonumber\\
& & \times \exp\{-\textmd{tr}(\textbf{R}_s^{-1}\textbf{T}_1) \}\bigg\}^{K+1},\nonumber
\end{eqnarray}
where $\textbf{T}_1$ is the total covariance constructed using both the training and the test data. It is given by
\begin{eqnarray}
\textbf{T}_1=\frac{K}{K+1}\textbf{S}_w+\frac{1}{K+1}\frac{1}{\sigma^2}\textbf{R}_s\textbf{R}^{-1}(\textbf{y}-\textbf{C}\bs{\beta})(\textbf{y}-\textbf{C}\bs{\beta})^\dagger,
\end{eqnarray}
where $\textbf{S}_w$ is the weighted sample covariance which assigns different weights to different secondary samples, i.e.,
\begin{eqnarray}\label{sw}
\textbf{S}_w=\frac{1}{K}\sum_{j=1}^{J}\sum_{k_j=1}^{K_j}\frac{1}{\sigma_j^2}\textbf{n}_{kj}\textbf{n}_{kj}^\dagger.
\end{eqnarray}
Using a similar approach under $\mathcal{H}_0$
\begin{eqnarray}
&&f_0\left(\textbf{y},\textbf{n}_1,\dots,\textbf{n}_K\right)\nonumber\\
&&=\bigg\{\frac{\sigma_1^{-2NK_1/K+1}\cdots\sigma_J^{-2NK_J/K+1}}{\pi^N \textmd{det}(\textbf{R})^{1/K+1}\textmd{det}(\textbf{R}_s)^{K/K+1}\sigma^{2N/K+1}}\nonumber\\
& & \times \exp\{-\textmd{tr}(\textbf{R}_s^{-1}\textbf{T}_0) \}\bigg\}^{K+1},\nonumber
\end{eqnarray}
where
\begin{eqnarray}
\textbf{T}_0=\frac{K}{K+1}\textbf{S}_w+\frac{1}{K+1}\frac{1}{\sigma^2}\textbf{R}_s\textbf{R}^{-1}(\textbf{y}-\textbf{B}\bs{\phi})(\textbf{y}-\textbf{B}\bs{\phi})^\dagger.
\end{eqnarray}
In the constrained GLRT using an ad-hoc approach, and considering the $\textbf{R}_s$ to be known in the first stage, the aim is to construct the following likelihood ratio test
\begin{eqnarray}\label{main1}
\ell_{GLRT}(\textbf{y})&=& \frac{\sup_{\bs{\beta},\sigma^2,\sigma^2_1,\cdots,\sigma^2_J,\textbf{R}} f_1\left(\textbf{y},\textbf{n}_1,\dots,\textbf{n}_K\right)}{\sup_{\bs{\phi},\sigma^2,\sigma^2_1,\cdots,\sigma^2_J,\textbf{R}} f_0\left(\textbf{y},\textbf{n}_1,\dots,\textbf{n}_K\right)}.\nonumber\\
&&\text{s.t.}\quad \|\textbf{R}-\textbf{R}_s\|_F^2\leq\epsilon\nonumber\\
&& \quad \quad\|\textbf{R}_s\|_F^2=\|\textbf{R}\|_F^2=1.
\end{eqnarray}
In (\ref{main1}) the ratio is the likelihood ratio test while the first constrain enforces the covariance matrices to be in the vicinity of each other with respect to Frobenius norm. In order to make the solutions of the base covariance matrices unique, i.e., $\textbf{R}$ and $\textbf{R}_s$ to be unique, we consider a second constraint to make their norm fixed. Their scales can be compensated by $\sigma$, instead.
We start with the denominator and try to first estimate $\bs{\phi}$ under $\mathcal{H}_0$. After using tedious but simple linear algebra, it can be shown that
\begin{eqnarray}\label{phi}
\hat{\bs{\phi}}=\left(\bar{\textbf{B}}^\dagger\bar{\textbf{B}}\right)^{-1}\bar{\textbf{B}}^\dagger\bar{\textbf{y}},
\end{eqnarray}
where $\bar{\textbf{B}}=\textbf{R}^{-\frac{1}{2}}\textbf{B}$ and $\bar{\textbf{y}}=\textbf{R}^{-\frac{1}{2}}\textbf{y}$. therefore, the denominator of (\ref{main1}) after one step maximization can be recast as
\begin{eqnarray}\label{1step}
&&\bigg\{\frac{\sigma_1^{-2NK_1/K+1}\cdots\sigma_J^{-2NK_J/K+1}}{\pi^N \textmd{det}(\textbf{R})^{1/K+1}\textmd{det}(\textbf{R}_s)^{K/K+1}\sigma^{2N/K+1}}\bigg\}^{K+1}\nonumber\\
&& \times \exp\{-\textmd{tr}({K}\textbf{R}_s^{-1}\textbf{S}_w)-\textmd{tr}(\frac{1}{\sigma^2}\bar{\textbf{y}}^\dagger\textbf{P}_{\bar{\textbf{B}}^\perp}\bar{\textbf{y}}) \}.
\end{eqnarray}
Now one should estimate $\sigma^2$ under $\mathcal{H}_0$. Starting with maximizing (\ref{1step}) with respect to $\sigma^2$, it can be shown that it is equivalent to following minimization 
\begin{eqnarray}
\hat{\sigma}^2&=&\arg\min_{\sigma^2}\frac{1}{\sigma^2}(\bar{\textbf{y}}^\dagger\textbf{P}_{\bar{\textbf{B}}^\perp}\bar{\textbf{y}})+N \log(\sigma^2),
\end{eqnarray}
which gives
\begin{eqnarray}
\hat{\sigma}^2&=&(\bar{\textbf{y}}^\dagger\textbf{P}_{\bar{\textbf{B}}^\perp}\bar{\textbf{y}})/N.
\end{eqnarray}
Since above estimator is independent of $\sigma^2_j, j=1,\cdots,J$ the maximization with respect to $\sigma^2_j, j=1,\cdots,J$ can be achieved by
\begin{eqnarray}
\hat{\sigma}_j^2&=&\arg\min_{\sigma_j^2}\textmd{tr}({K}\textbf{R}_s^{-1}\textbf{S}_w)+NK_j \log(\sigma_j^2)\nonumber\\
&=&\arg\min_{\sigma_j^2}\textmd{tr}(\textbf{R}_s^{-1}\sum_{j=1}^{J}\sum_{k_j=1}^{K_j}\frac{1}{\sigma_j^2}\textbf{n}_{k_j}\textbf{n}_{k_j}^\dagger)+NK_j \log(\sigma_j^2)\nonumber\\
&=&\arg\min_{\sigma_j^2}\frac{1}{\sigma_j^2}\textmd{tr}(\textbf{R}_s^{-1}\sum_{k_j=1}^{K_j}\textbf{n}_{k_j}\textbf{n}_{k_j}^\dagger)+NK_j \log(\sigma_j^2).\nonumber
\end{eqnarray}
This gives
\begin{eqnarray}\label{sigmatrain}
\hat{\sigma}_j^2&=&\frac{1}{NK_j}\sum_{k_j=1}^{K_j}\textbf{n}_{k_j}^\dagger\textbf{R}_s^{-1}\textbf{n}_{k_j}, \text{  }j=1,\cdots,J.
\end{eqnarray}
Now, we maximize the denominator with respect to $\textbf{R}$. Since there is a constraint for the covariance matrix to be in the proximity of the $\textbf{R}_s$, the closed form solution is not available and we use iterative optimization algorithm to converge to a proper solution. The optimization problem for estimating $\textbf{R}$ is approximated by
\begin{eqnarray}
\hat{\textbf{R}}&=&\arg\min_{\textbf{R}}\log\textmd{det}(\textbf{R})+ (K+1)\textmd{tr}(\textbf{R}_s^{-1}\textbf{T}_0) \nonumber\\
&&\text{s.t.}\quad \|\textbf{R}-\textbf{R}_s\|_F^2\leq\epsilon\nonumber\\
&& \quad \quad\|\textbf{R}\|_F^2\leq1\\
&=&\arg\min_{\textbf{R}}\log\textmd{det}(\textbf{R})+\frac{1}{\sigma^2}\textmd{tr}(\textbf{R}^{-1}(\textbf{y}-\textbf{B}\bs{\phi})\nonumber\\
&&(\textbf{y}-\textbf{B}\bs{\phi})^\dagger) \nonumber\\
&&\text{s.t.}\quad \|\textbf{R}-\textbf{R}_s\|_F^2-\epsilon\leq 0 \nonumber\\
&& \quad \quad\|\textbf{R}\|_F^2\leq1
\end{eqnarray}
In order to be able to use the Alternating Direction Methods of Multipliers (ADMM) \cite{giesen2016distributed}, we convert this constrained loss into two separable loss functions with affine equality constraint and convex inequality constraints, given by

\begin{eqnarray}\label{cons}
\hat{\textbf{R}},\hat{\textbf{Z}}&=&\arg\min_{\textbf{R},\textbf{Z}}\log\textmd{det}(\textbf{R})+\frac{1}{\sigma^2}\textmd{tr}(\textbf{Z}^{-1}(\textbf{y}-\textbf{B}\bs{\phi})\nonumber\\
&&(\textbf{y}-\textbf{B}\bs{\phi})^\dagger) \nonumber\\
&&\text{s.t.}\quad \|\textbf{R}-\textbf{R}_s\|_F^2-\epsilon\leq 0 \nonumber\\
&& \quad \quad\|\textbf{R}\|_F^2\leq1\nonumber\\
&&\quad \quad \textbf{R}-\textbf{Z}=\textbf{0},
\end{eqnarray}
where $\epsilon$ is the hyperparameter controlling the level of homogeneity. Using the Augmented Lagrangian technique with parameter $\rho>0$, our new problem is to minimize $L_\rho(\textbf{R},\textbf{Z},\textbf{U},\gamma,\lambda)$, i.e.,
\begin{eqnarray}
&&L_\rho(\textbf{R},\textbf{Z},\textbf{U},\gamma,\lambda)=\log\textmd{det}(\textbf{R})+\frac{1}{\sigma^2}\textmd{tr}(\textbf{Z}^{-1}(\textbf{y}-\textbf{B}\bs{\phi})\nonumber \\
&&(\textbf{y}-\textbf{B}\bs{\phi})^\dagger)+\rho/2\|\text{max}(0, \|\textbf{R}-\textbf{R}_s\|_F^2-\epsilon)^2\|^2\nonumber\\
&&+\lambda\text{max}(0, \|\textbf{R}-\textbf{R}_s\|_F^2-\epsilon)^2+\rho/2\|\text{max}(0, \|\textbf{R}\|_F^2-1)^2\|^2\nonumber\\
&&+\gamma\text{max}(0, \|\textbf{R}\|_F^2-1)^2\nonumber\\
&&  +\rho/2 \|\textbf{R}-\textbf{Z}\|_F^2+\text{tr}(\textbf{U}(\textbf{R}-\textbf{Z})),
\end{eqnarray}
where $\textbf{U}$, $\gamma$ and $\lambda$ are the Lagrangian multipliers. Now, using the ADMM updating rules, the iterative estimators are
\begin{eqnarray}\label{Radmm}
\textbf{R}^{t+1}&=& \textbf{R}^{t}-\eta \nabla_{\textbf{R}}{L_\rho(\textbf{R}^t,\textbf{Z}^t,\textbf{U}^t,\gamma^t,\lambda^t)}\nonumber \\ &=& \textbf{R}^{t}-\eta\big\{ \textbf{R}^{-t}+4\rho HS(\|\textbf{R}^t-\textbf{R}_s\|_F^2-\epsilon)\nonumber\\
&&\times \text{max}(0, \|\textbf{R}^t-\textbf{R}_s\|_F^2-\epsilon)^3(\textbf{R}^{t}-\textbf{R}_s)+4\lambda^{t} \nonumber\\
&&\times HS(\|\textbf{R}^t-\textbf{R}_s\|_F^2-\epsilon)\text{max}(0, \|\textbf{R}^t-\textbf{R}_s\|_F^2-\epsilon)\nonumber\\
&&\times(\textbf{R}^{t}-\textbf{R}_s)+4\rho HS(\|\textbf{R}^t\|_F^2-1)\nonumber\\
&&\times \text{max}(0, \|\textbf{R}^t\|_F^2-1)^3\textbf{R}^{t}+4\gamma^{t} HS(\|\textbf{R}^t\|_F^2-1) \nonumber\\
&&\times \text{max}(0, \|\textbf{R}^t\|_F^2-1)\textbf{R}^{t}+\textbf{U}^\top+\rho(\textbf{R}^{t}-\textbf{Z})\big\}\nonumber,\\
\end{eqnarray}

\begin{eqnarray}
\textbf{Z}^{t+1}&=& \textbf{Z}^{t}-\eta \nabla_{\textbf{Z}}{L_\rho(\textbf{R}^{t+1},\textbf{Z}^t,\textbf{U}^t,\gamma^t,\lambda^t)} \\ &=& \textbf{Z}^{t}-\eta\big\{-\frac{1}{\sigma^2}\textbf{Z}^{-t}(\textbf{y}-\textbf{B}\bs{\phi})(\textbf{y}-\textbf{B}\bs{\phi})^\dagger\textbf{Z}^{-t}\nonumber \\ && +\rho(\textbf{Z}^{t}-\textbf{R}^{t+1}) -{\textbf{U}^{t}}^\top\big\},\nonumber
\end{eqnarray}
\begin{figure}[!t] 
	\centering
	\includegraphics[width=0.9\linewidth]{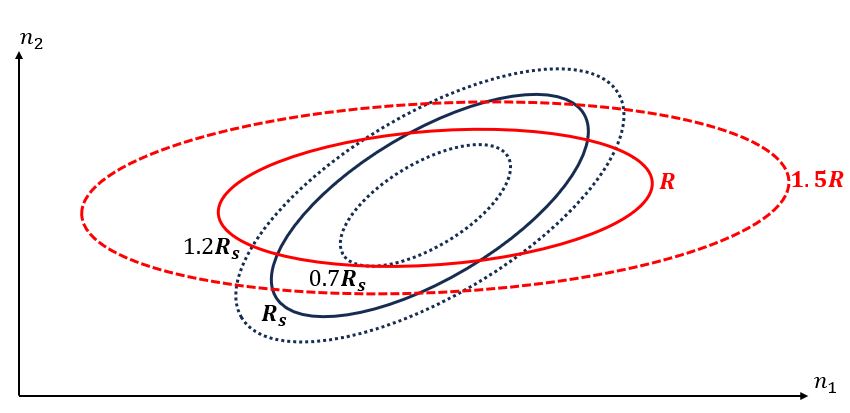}
	\caption{Visualizing the covariance matrices in the training (black) set and test set (red) in two dimensional datasets. Dashed lines show multiple scaled version of the base covariance matrices (solid lines). The proposed framework enforces the base covariance matrices be almost the same, while the observed covariance matrices are different in scales. }
	\label{fig2}
\end{figure}
\begin{figure*}[!t] 
	\centering
	\includegraphics[width=0.48\linewidth]{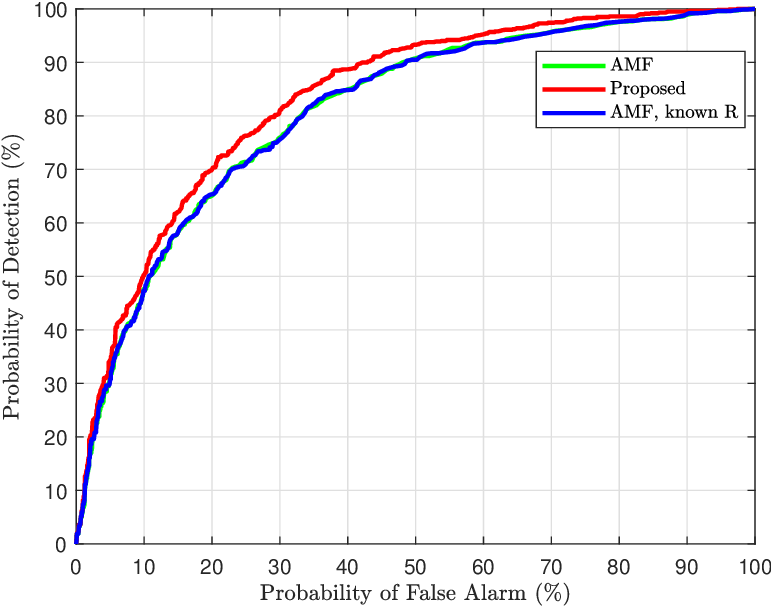}
	\includegraphics[width=0.48\linewidth]{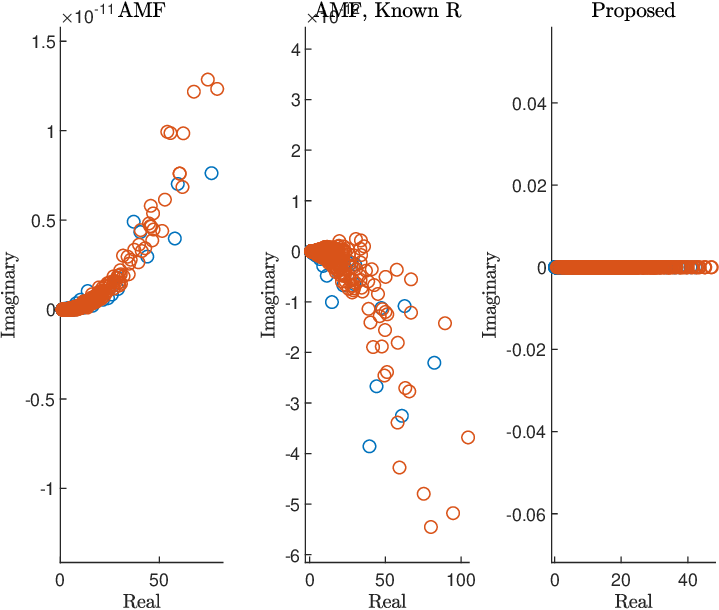}
	\caption{Performance visualizing of the proposed detector (ROC in the left, output histogram in the right) in comparison with AMF with estimated covariance and know covariance.  }
	\label{fig3}
\end{figure*}
\begin{eqnarray}
\textbf{U}^{t+1}&=& \textbf{U}^{t}+\rho \big\{\textbf{R}^{t+1}-\textbf{Z}^{t+1}\big\},
\end{eqnarray}
\begin{eqnarray}
\gamma^{t+1}&=& \gamma^{t}+\rho \big\{\text{max}(0, \|\textbf{R}^{t+1}\|_F^2-1)^2\big\},
\end{eqnarray}
\begin{eqnarray}
\lambda^{t+1}&=& \lambda^{t}+\rho \big\{\text{max}(0, \|\textbf{R}^{t+1}-\textbf{R}_s\|_F^2-\epsilon)^2\big\}.
\end{eqnarray}
When $t\rightarrow \infty$, the estimator $\textbf{R}$ converges to a minimum, where we call it $\textbf{R}_{\infty}$. Similarly, the MLE of unknown parameters under $\mathcal{H}_1$ are given by
\begin{eqnarray}\label{beta}
\hat{\bs{\beta}}&=&\left(\bar{\textbf{C}}^\dagger\bar{\textbf{C}}\right)^{-1}\bar{\textbf{C}}^\dagger\bar{\textbf{y}},\nonumber\\
\hat{\sigma}^2&=&(\bar{\textbf{y}}^\dagger\textbf{P}_{\bar{\textbf{C}}^\perp}\bar{\textbf{y}})/N ,\nonumber\\
\hat{\sigma}_j^2&=&\frac{1}{NK_j}\sum_{k_j=1}^{K_j}\textbf{n}_{k_j}^\dagger\textbf{R}_s^{-1}\textbf{n}_{k_j}, \text{  }j=1,\cdots,J,\nonumber
\end{eqnarray} 
where $\bar{\textbf{C}}=\textbf{R}^{-\frac{1}{2}}\textbf{C}$. $\textbf{R}$ can be estimated the same approach described for $\textbf{R}$ in null hypothesis (\ref{Radmm}), where the matrix $\textbf{B}$ is replaced with matrix $\textbf{C}$. Now with substitution of above estimators in (\ref{main1}), it returns the likelihood ratio for a decision maker. This detector still assumes a known covariance matrix $\textbf{R}_s$, which is unknown in reality. Thus, in the second step of design process, we replace this matrix with an appropriate estimator. To do so, we form the joint likelihood function in the secondary data, given by

\begin{eqnarray}
&&f_s\left(\textbf{n}_1,\dots,\textbf{n}_K\right)\nonumber\\
&&=\left\{\frac{\pi^{-N} \textmd{det}(\textbf{R}_s)^{-1}}{\sigma_1^{2NK_1/K}\cdots \sigma_J^{2NK_J/K}}\right\}^{K}\times\prod_{k_1=1}^{K_1}\nonumber\\
&&\exp\{-\frac{1}{\sigma_1^2}\textbf{n}_{k_1}^\dagger \textbf{R}_s^{-1} \textbf{n}_{k_1}\}\cdots\prod_{k_J=1}^{K_J}\exp\{-\frac{1}{\sigma_J^2}\textbf{n}_{k_J}^\dagger \textbf{R}_s^{-1} \textbf{n}_{k_J}\}.\nonumber
\end{eqnarray}
The ML estimation of the covariance can be found by solving
\begin{eqnarray}
\hat{\textbf{R}}_s&=&\arg\max_{\textbf{R}_s}\log\{f_s\left(\textbf{n}_1,\dots,\textbf{n}_K\right)\}\nonumber \\
&=&\arg\min_{\textbf{R}_s} K\log\{\textmd{det}(\textbf{R}_s)\}+\textmd{tr}(K\textbf{R}_s^{-1} \textbf{S}_w)\nonumber,
\end{eqnarray}
The solution of above problem is $\hat{\textbf{R}}_s=\textbf{S}_w$ which is the weighted sample covariance matrix. However, this would be problematic since, now estimation of $\sigma_j^2$ relies on $\textbf{R}_s$ and vice versa, while estimator of $\textbf{R}$ also relies on  $\textbf{R}_s$. Therefore, we use a alternation technique between (\ref{sw}) and (\ref{sigmatrain}) together with the normalization of $\textbf{R}_s$ to estimate these parameters. Figure \ref{fig2} illustrates a visualization of the assumed setup for a two-dimensional observation case. The goal is to estimate the base $\textbf{R}$ and $\textbf{R}_s$ such that they are approximately similar, and meanwhile they can maximize the likelihood function. Compared to the AMF introduced in (\ref{AMF}), this detector is designed for a more general case that the noise samples in the secondary data can be drawn from different but still circularly symmetric complex Gaussian distributions. The final form of detector, after substitution of all estimated parameters in (\ref{main1}) is given by

\begin{eqnarray}\label{final}
\ell_{GLRT}(\textbf{y})&=&\mathcal{R}\bigg\{ \frac{(\bar{\textbf{y}}^\dagger\textbf{P}_{\bar{\textbf{B}}^\perp}\bar{\textbf{y}})^{2N}\textmd{det}(\textbf{R}_{\infty|H_0})}{(\bar{\textbf{y}}^\dagger\textbf{P}_{\bar{\textbf{C}}^\perp}\bar{\textbf{y}})^{2N}\textmd{det}(\textbf{R}_{\infty|H_1})}\bigg\},
\end{eqnarray}
where $\textbf{R}_{\infty|H_i}$ indicates the covariance estimation under $i^{th}$ hypothesis and $\mathcal{R}\{.\}$ takes the real part. As visible, the above detector is a general form of AMF. In the case of homogeneous environment, under some setups,  $\textbf{R}_{|H_i}=\textbf{R}_s$ and the AMF that was introduced in (\ref{AMF}) will be obtained.

\begin{figure*}[!t] 
	\centering
	\includegraphics[width=0.32\linewidth]{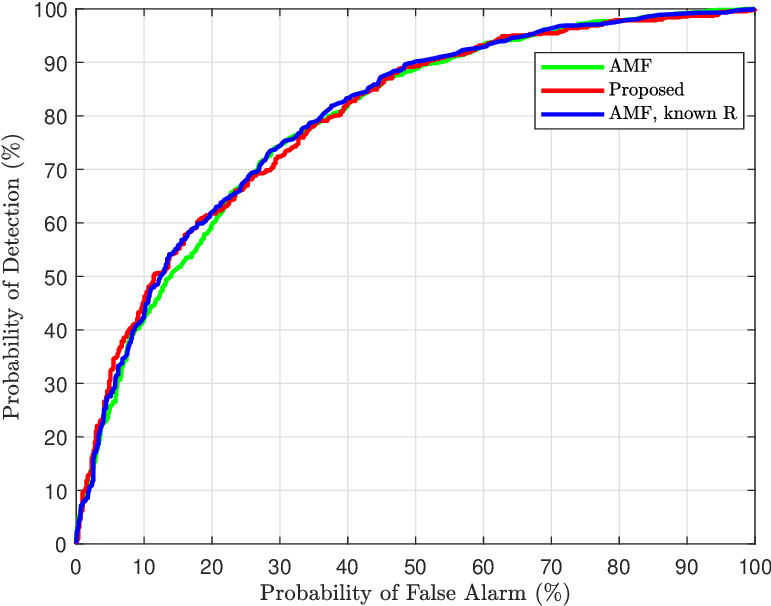}
	\includegraphics[width=0.32\linewidth]{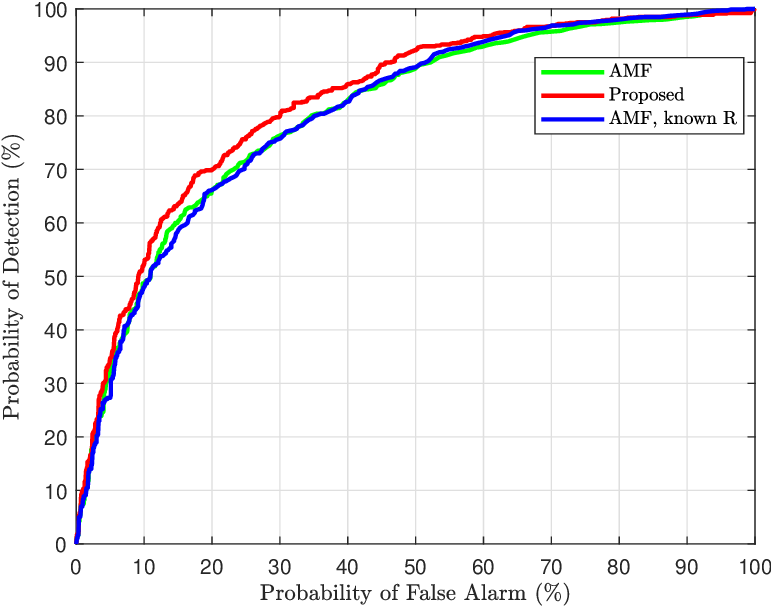}
	\includegraphics[width=0.32\linewidth]{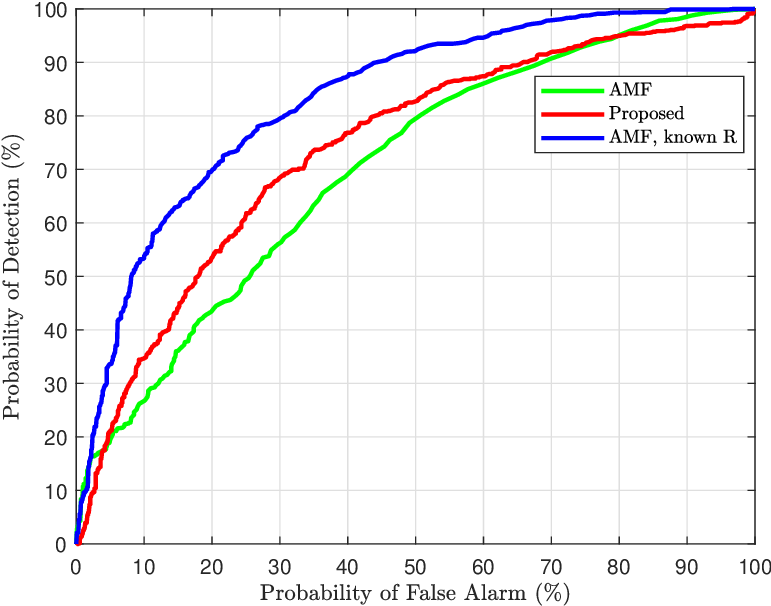}
	\caption{Performance visualizing of the proposed detector (ROC) in comparison with AMF in partially homogenous environment  (left), non-stationary partially homogenous environment (middle), and  heterogeneous environment (right).  }
	\label{fig4}
\end{figure*}
\section{Simulation Results}
\label{sec:results}

To test the proposed detector, we set the setup as described below and then gradually make more complicated scenarios to show the superiority of the proposed detector in dynamical environment. \\
\textbf{Setup:} We set $N=5$, $p=2$, $t=1$, and maximum number of iteration to $2000$. The signal to noise ratio (SNR) is defined as $\text{SNR}=\text{10 log}(\boldsymbol{\theta}^\dag\textbf{H}^\dag(\sigma^2\textbf{R})^{-1}\textbf{H}\boldsymbol{\theta}),$ and the subspace matrix $\textbf{H}$ is chosen to be
\begin{equation}
\textbf{H}=[\textbf{h}_1, \textbf{h}_2, \cdots, \textbf{h}_p],
\end{equation}
where
\begin{equation}
\textbf{h}_i=1/\sqrt{N}[1, e^{-j2\pi f_i}, \cdots, e^{-j2\pi f_i(N-1)}]^\top,
\end{equation}
and $f_i=i\times0.05+0.05, i=1,\cdots,p$. The subspace matrix $\textbf{B}$ is chosen in a similar way where
\begin{equation}
\textbf{b}_i=1/\sqrt{N}[1, e^{-j2\pi g_i}, \cdots, e^{-j2\pi g_i(N-1)}]^\top,
\end{equation}
and $g_i=-i\times0.025+0.025, i=1,\cdots,t$. Now we examine the proposed method in different setups.\\
\textbf{Homogeneous Environment (HE) with large $K$:} When K is large (500 in this experiment), the covariance estimate is a good approximation of the actual covariance. Setting SNR into $8dB$, we set $\epsilon=0$ to ensure we are dealing with a homogeneous environment, $\rho=2$ to ensure smooth convergence of $\textbf{Z}$ to $\textbf{R}$ and smooth satisfaction of other constraints in (\ref{cons}), $\eta=0.0001$ and all the training and test $\sigma^2$ to $5$, where the actual base covariance matrices for both training and test is the same (no dynamic). This covariance matrix is defined as $\textbf{R}_s=\textbf{R}=0.44\textbf{I}$. This means there is no correlation among dimensions. Figure \ref{fig3} (left) illustrates the performance of the proposed method in comparison with AMF where the covariance matrix was estimated using the secondary dataset and AMF where the actual covariance matrix was used. As shown, the proposed method is able to perform as good as AMF (slightly better) since parameter $\epsilon$ enforced the covariance matrix to be almost the same as estimated $\textbf{R}_s$ and meanwhile the test data information leaked to covariance estimation. Figure \ref{fig3} (right) illustrates the distribution of these three test statistics for $\mathcal{H}_1$ (red circles) and $\mathcal{H}_0$ (blue circles). In this experiment, the performance was reported for $2000$ test samples.\\
\textbf{Partially Homogenous Environment (PHE) with medium $K$:} For all the training samples in the secondary set, we use $\sigma_k^2=5$ and for the test samples we use $\sigma^2=20$, with the same base covariance matrices where $\textbf{R}_{s}=\textbf{R}=0.44\textbf{I}$. We also fix SNR into $8dB$ as before with $K=40$, and 2000 samples to computer the ROC and output distributions. Note that AMF is still expected to perform well, since its design assumptions is not violated yet. We set $\epsilon=0$ to ensure consistency of the covariance between train and test with a large penalty, $\rho=2$, and $\eta=0.0001$. As shown in Figure \ref{fig4} (left), the proposed method closely follows the AMF, since using the design approach and fixed parameters, the proposed method tends to be a replica of AMF.\\
\textbf{Non-Stationary Partially Homogenous Environment (PHE) with large $K$:} In this setup, while the base covariance matrices are the same, still  $\textbf{R}_{s}=\textbf{R}=0.44\textbf{I}$, half of the secondary set are generated by $\mathcal{C}\mathcal{N}(\textbf{0},5\textbf{R}_s)$ and the other half are generated by $\mathcal{C}\mathcal{N}(\textbf{0},15\textbf{R}_s)$ and the test sample noise is drawn from $\mathcal{C}\mathcal{N}(\textbf{0},30\textbf{R}_s)$. Using the same parameters as before, Figure \ref{fig4} (middle), shows the ROC for such scenario. Interestingly, although samples in secondary set are generated using different scales of base covariance $\textbf{R}_s$, but since the test covariance is still the scaled version of the same base covariance, the result in AMF is robust to such changes and almost the same ROC pattern as observed as the case of PHE. However, due to additional information in the test data, ROC of the proposed method is slightly better than its counterparts.\\
\textbf{Heterogeneous Environment with large $K$:} In this more realistic scenario, while we use  $\textbf{R}_{s}=0.44\textbf{I}$, but for the test data, we use $\textbf{R}_{i,j}=(\textbf{R}_{s_{i,j}}+\alpha 0.95^{|i-j|})$ followed by a normalization step with $8dB$ SNR. We use $\alpha=2$, $\sigma_1^2=5$, $\sigma_2^2=15$, $\sigma^2=30$ and $\epsilon=0.2$. Figure \ref{fig4} (right) illustrates the ROC performance of the proposed method compared to AMF variants and the proposed method outperforms AMF where the covariance matrix was estimated using PHE assumption. However, there is a large gap in performance compared to the case where the covariance is known (solid blue curve) which can be only compensated with more observations in test stage (distributed target scenario).

\section{Discussion}
\label{discussion}

\begin{figure}[!t] 
	\centering
	\includegraphics[width=1\linewidth]{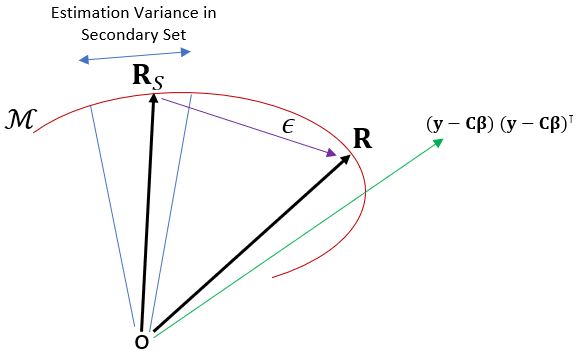}
	\caption{Geometry of the estimated covariance over bounded norm manifold in secondary set ($\hat{\textbf{R}}_s$) and its deviation from the noise covariance at testing stage ($\textbf{R}$). The single point-based covariance estimation is shown in green, i.e., $(\textbf{y}-\textbf{C}\boldsymbol{\beta})(\textbf{y}-\textbf{C}\boldsymbol{\beta})^\dagger$. }
	\label{fig5}
\end{figure}

Single-step GLRT based techniques offer to use the MLE of the form $\frac{K}{K+1}\textbf{S}+\frac{1}{\sigma^2}(\textbf{y}-\textbf{C}\boldsymbol{\beta})(\textbf{y}-\textbf{C}\boldsymbol{\beta})^\dagger$ for the test covariance and they consider the new test sample (possibly with different covariance structure) in their detector design  which help to attain better performance. However, for large K value, this improvement over two-step detectors become negligible. On the other hand, two-step detectors never take into account the test samples in their covariance estimation step, which makes them intolerable with respect to change in covariance structure between training and test datasets (e.g., AMF). In our proposed technique, even though the design is based on two-step procedure, however, we aimed to find the best matched covariance to the test sample with constraints to be similar to training covariance with defining a measure of vicinity. Regardless of the number of samples in secondary set (K), the proposed method always seeks for the best matches to the test sample to capture the dynamical changes (possibly sever changes) between two datasets (see Figure \ref{fig5}). Without the considered constraints, the solution of the covariance structure would be $\frac{1}{\sigma^2}(\textbf{y}-\textbf{B}\bs{\phi})(\textbf{y}-\textbf{B}\bs{\phi})^\dagger$, which is rank 1 and a non-invertible matrix.  This can bee seen as the main difference between the proposed method and the other counterparts in the literature.
\section{Conclusion}
\label{sec:conclusion}
%------------------------------------------------------------------
In this paper, we addressed the problem of subspace detection in the case where the  noise covariance between the training and testing stages might significantly differ from each other. This setup is known as heterogeneous environment and  we have shown that, although th design process is two-step, in contrast to other two-step detectors, the proposed technique can fully track the changes in covariance structures and this leads to a better detection rate. Simulation results in different setups revealed the superiority of the detector and showed that the proposed detector unifies and also extends the existing two-step detectors. In future works, the same idea can be applied to one-step detectors to leverage their strength in using additional information in the test data. The results can be also extended into structured covariance matrices with distributed target scenarios. 
\bibliographystyle{IEEEbib}
\bibliography{References}

\begin{IEEEbiography}[{\includegraphics[width=1.1in,height=1.9in,clip,keepaspectratio]{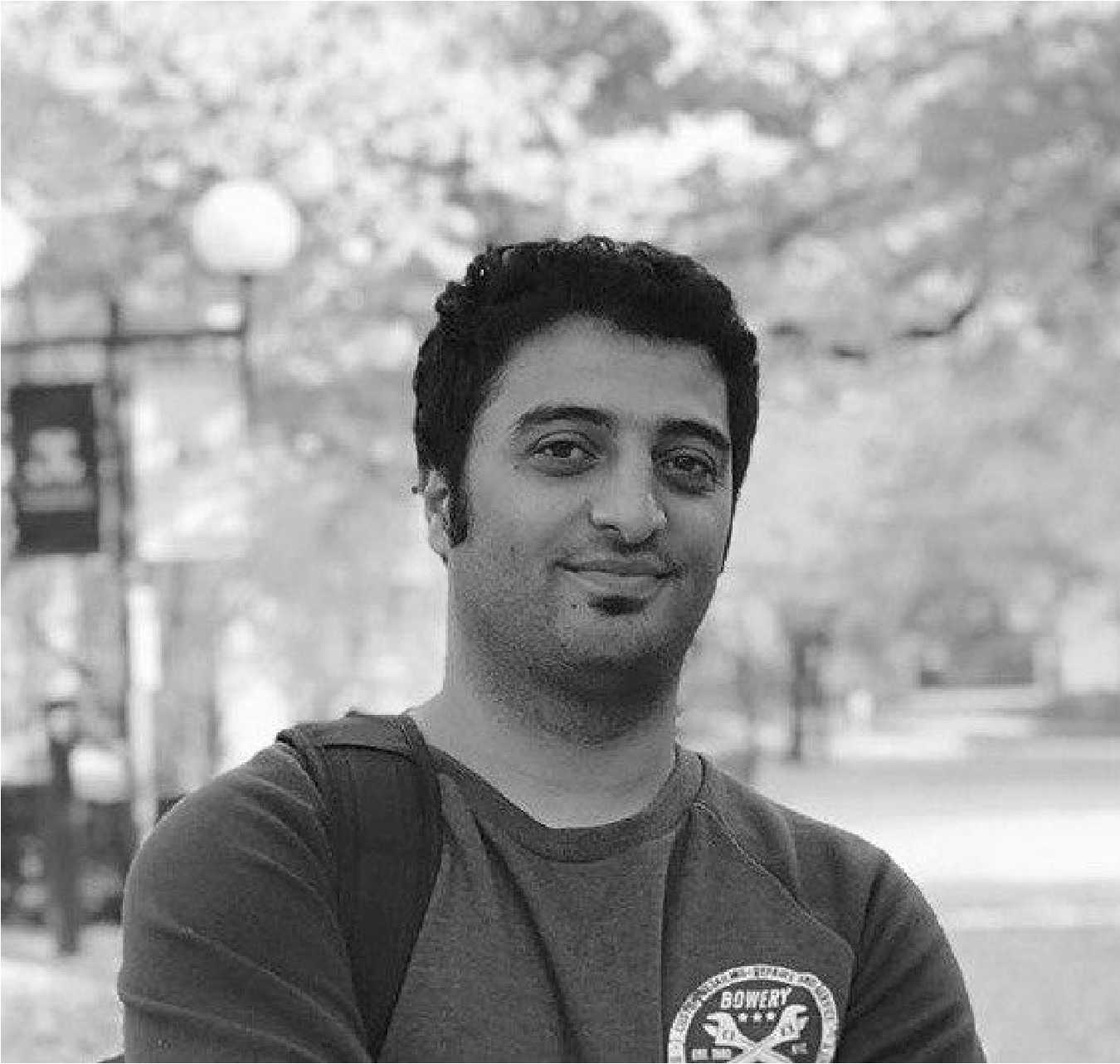}}]{Aref Miri Rekavandi}(Member, IEEE) received the B.Sc. degree in electrical engineering from KN Toosi University of Technology, Tehran, Iran, in 2013, M.Sc. degree in digital electronics engineering from Amirkabir University of Technology (Tehran Polytechnic), Tehran, Iran, in 2016, and Ph.D. degree in electrical and electronic engineering from The University of Melbourne, Melbourne, VIC, Australia, in 2021. He is currently
a Research Fellow at Faculty of Engineering and Information Technology, The University of Melbourne, Australia. Prior to this role, he was a postdoctoral researcher (2021-2023) at the School of Physics, Maths and Computing, The University of Western Australia, Australia. He was also a Visiting Researcher at NDCN, The University of Oxford and a Visiting Researcher at the School of Mathematics and Statistics, The University of Melbourne.  His research interests include statistical signal and image processing, detection theory, computer vision, medical imaging and pattern recognition.
\end{IEEEbiography}%

\end{document}